\documentclass[a4paper,fleqn,usenatbib]{mnras}

\usepackage[T1]{fontenc}
\usepackage{ae,aecompl}

\usepackage{graphicx}	% Including figure files
\usepackage{amsmath}	% Advanced maths commands
\usepackage{amssymb}	% Extra maths symbols

\newcommand{\jed}[1]{~{\rm #1}}

\title[Planetary candidate around giant star HD~175370]
{The discovery of a planetary candidate around the evolved low-mass \textit{Kepler} giant star HD~175370
\thanks{Based on observations made with NASA\textquoteright s Discovery mission
\textit{Kepler} and with the HERMES spectrograph,
installed at the Mercator telescope, operated on the island of La Palma by the Flemish Community,
at the Spanish Observatorio del Roque de los Muchachos of the Instituto de
Astrof\'isica de Canarias and supported by the Research Foundation - Flanders (FWO), Belgium, the
Research Council of KU Leuven, Belgium, the Fonds National de la Recherche Scientific 
(F.R.S.-FNRS), Belgium, the Royal Observatory of Belgium, the Observatoire de Gen\`eve, Switzerland
and the Th\"uringer Landessternwarte Tautenburg, Germany.}}

\author[M.~Hrudkov\'{a} et al.]{M.~Hrudkov\'{a}$^{1}$\thanks{E-mail: mh@ing.iac.es}, 
A.~Hatzes$^{2}$, R.~Karjalainen$^{1}$, H.~Lehmann$^{2}$, S.~Hekker$^{3,4}$, 
\newauthor
M.~Hartmann$^{2}$, A.~Tkachenko$^{5}$, S.~Prins$^{5}$, H.~Van~Winckel$^{5}$, R.~De~Nutte$^{6}$,
\newauthor
L.~Dumortier$^{6}$, Y.~Fr\'emat$^{6}$, H.~Hensberge$^{6}$, A.~Jorissen$^{7}$, P.~Lampens$^{6}$, 
\newauthor
M.~Laverick$^{5}$, R.~Lombaert$^{8,5}$, P.~I.~P\'apics$^{5}$, G.~Raskin$^{5}$, \'A.~S\'odor$^{6,9}$, 
\newauthor
A.~Thoul$^{10,11}$, S.~Van~Eck$^{7}$, C.~Waelkens$^{5}$\\
$^{1}$Isaac Newton Group of Telescopes, Apartado de Correos 321, Santa Cruz
de La Palma, E-38700, Spain\\
$^{2}$Th\"uringer Landessternwarte Tautenburg, Sternwarte 5, Tautenburg,
D-07778, Germany\\
$^{3}$Max-Planck-Institut f\"ur Sonnensystemforschung, Justus-von-Liebig-Weg
3, G\"ottingen, D-37077, Germany\\
$^{4}$Stellar Astrophysics Centre, Department of Physics and Astronomy,
Aarhus University, Ny Munkegade 120, Aarhus C, DK-8000, Denmark\\
$^{5}$Instituut voor Sterrenkunde, KU Leuven, Celestijnenlaan 200D bus 2401,
Leuven, 3001, Belgium\\
$^{6}$Royal Observatory of Belgium, 3 Avenue Circulaire, Brussel, 1180,
Belgium\\
$^{7}$Institut d\textquoteright Astronomie et d\textquoteright
Astrophysique, Universit\'e Libre de Bruxelles, CP 226,
Boulevard du Triomphe, Bruxelles, 1050, Belgium\\
$^{8}$Department of Earth and Space Sciences, Chalmers University of
Technology, Onsala Space Observatory, 439 92 Onsala, Sweden\\
$^{9}$Konkoly Observatory, Research Centre for Astronomy and Earth Sciences,
Hungarian Academy of Sciences, Budapest, 1121, Hungary\\
$^{10}$Institut d\textquoteright Astrophysique et de G\'eophysique,
Universit\'e de Li\`ege, 17 All\'ee
du 6 Ao$\hat{u}$t, 4000 Li\`ege, Belgium\\
$^{11}$Kavli Institute for Theoretical Physics, Kohn Hall, University of
California, Santa Barbara CA 93106-4030, USA
}

\date{Accepted XXX. Received YYY; in original form ZZZ}

\pubyear{2016}

\begin{document}
\label{firstpage}
\pagerange{\pageref{firstpage}--\pageref{lastpage}}
\maketitle

\begin{abstract}
We report on the discovery of a planetary companion candidate with a minimum mass
$M\,\sin i=4.6\pm 1.0\,M_{\rm{Jupiter}}$ orbiting the K2$\,$III giant star
HD~175370 (KIC~007940959). This star was a target in our program to search for planets
around a sample of 95 giant stars observed with \textit{Kepler}.
This detection was made possible using precise stellar radial velocity 
measurements of HD~175370 taken over five years and four months  
using the coud\'e echelle spectrograph of the 2-m Alfred Jensch Telescope
and the fibre-fed echelle spectrograph HERMES of the 1.2-m Mercator Telescope.
Our radial velocity measurements reveal a
periodic ($349.5\pm 4.5$ days) variation with a semi-amplitude
$K=133\pm 25\,\rm{m\,s^{-1}}$, superimposed on a long-term trend. A low-mass
stellar companion with an orbital period of $\sim 88$ years in a highly eccentric orbit
and a planet in a Keplerian orbit with an eccentricity
$e=0.22$ are the most
plausible explanation of the radial velocity variations. However, we cannot
exclude the existence of stellar envelope pulsations as a cause for the low-amplitude radial
velocity variations and only future continued monitoring of this system may
answer this uncertainty. From \textit{Kepler} photometry we find that HD~175370 is most likely a low-mass
red-giant branch or asymptotic-giant branch star.
\end{abstract}

\begin{keywords}
methods: observational -- methods: data analysis -- techniques: spectroscopic
-- techniques: radial velocities -- planetary systems -- stars: individual: HD~175370 (KIC~007940959).
\end{keywords}

%%%%%%%%%%%%%%%%%%%%%%%%%%%%%%%%%%%%%%%%%%%%%%%%%%

%%%%%%%%%%%%%%%%% BODY OF PAPER %%%%%%%%%%%%%%%%%%

\section{Introduction}\label{intro}

Planets around K-giant stars may provide us with clues on the dependence of
planet formation on stellar mass. The progenitors of K-giant stars are often
intermediate-mass main-sequence A-F stars. While on the main sequence
these stars are not amenable to precise radial velocity (RV) measurements.
There is a paucity of stellar lines due to
high effective temperatures and these are often broadened by rapid rotation.
Therefore one cannot easily achieve the RV precision needed for the detection of 
planetary companions.
On the other hand, when these stars evolve to giant stars they are cooler and have
slower rotation rates and a RV accuracy of a few$\,\,\rm{m\,s^{-1}}$ can
readily be achieved. The giant stars thus serve as proxies for planet searches 
around intermediate-mass (1.2--2 $M_{\odot}$) early-type main-sequence stars.

Since the discovery of the first exoplanet around K-giant stars
\citep{Hatzes93,Frink02}, over 90 exoplanets (3 per cent\footnote{The Extrasolar Planets Encyclopedia:
http://exoplanet.eu} of the total in September 2016) 
have been discovered orbiting giant stars. 
These planet-hosting giant stars are on average more massive than 
planet-hosting main-sequence 
stars. \citet{Johnson10} showed, based on an empirical correlation, 
that the frequency of giant planets increases with 
stellar mass to about 14 per cent for A-type stars. This is consistent with theoretical 
predictions of \citet{Kennedy08}, who concluded that the probability that a
given star has at least one gas giant increases linearly with stellar mass
from 0.4 to 3 $M_\odot$. Statistical analysis of microlensing
and transiting data reveals
that cool Neptunes and super-Earths are even more common than Jupiter-mass planets
\citep{Cassan12,Howard12}. 

Unlike for a main-sequence star where there is more or less a direct mapping
between effective temperature and stellar mass, it is more problematic 
to determine the stellar mass of a giant star. The evolutionary tracks
for stars covering a wide range of masses all converge to the
similar region of the H-R diagram. One way to obtain the stellar mass is to rely on evolutionary tracks.
However, they are not only model dependent, but they require accurate stellar parameters such
as the effective temperature and heavy element abundance. \citet{Lloyd11} argued that the masses of giant
stars in Doppler surveys were only in the range 1.0 -- 1.2 $M_\odot$ and thus were not
intermediate-mass stars. Clearly, we cannot disentangle the effect of stellar mass on the observed planet
properties if we cannot get a reliable measurement of the stellar mass.

The stellar mass can be derived from solar-like oscillations. 
The first firm discovery of solar-like oscillations in a giant star 
was made using RVs by \citet{Frandsen02}. However, it was only recently that
solar-like oscillations were unambiguously found in late-type
giant stars, owing to space-based photometric observations by the \textit{CoRoT} 
\citep{deRidder09} and the \textit{Kepler} \citep{Gilliland10} missions.
Solar-like p-mode oscillations of the same degree are equally spaced in frequency and the spacing 
is related to the square root of the mean stellar density
($\propto$$(M/R^{3})^{1/2}$), whereas
the frequency of maximum oscillation power is $\propto$ $M/(R^2
\sqrt{T_{\rm{eff}}})$ \citep{Kjeldsen95}. From these empirical relations we
can calculate both the stellar mass and radius in a more or less model
independent way.

The \textit{Kepler} Space Mission has been monitoring a sample of over 13,000
red-giant stars which can be used for asteroseismic studies. All stars
show stellar oscillations that have been analysed to determine their fundamental
stellar parameters \citep{Huber10,Stello13} and internal
stellar structure \citep{Bedding11}. 
Furthermore, an estimate of the stellar age can be obtained using stellar
models \citep{Lebreton14}. 

Giant stars observed with \textit{Kepler} represent a unique sample for planet searches as
a planetary detection would mean that we can determine
reliable stellar properties via asteroseismic analysis, characteristics not
well known for many other planet-hosting giant stars. For this reason, we 
started a planet-search program among \textit{Kepler} asteroseismic-giant
stars in 2010. We have distributed our targets over four different telescopes in order
to maximize the detection and to minimize the impact of telescope resources
at a single site. These telescopes include the 2-m telescope at Th\"uringer
Landessternwarte Tautenburg (TLS), Germany (29 stars, since 2010), the 1.2-m Mercator telescope, La
Palma, Spain (38 stars, since 2011), the 2.5-m Nordic Optical Telescope, La
Palma, Spain (12 stars, since 2012) and the 2.7-m telescope at McDonald Observatory,
Texas, US (33 stars, since 2012). In total our sample contains 95 giant stars, a
statistically significant number given an expected detection rate of
$\sim$15 per cent. We observe some targets in common at different sites, in order 
to check our measurements independently.

Until now, three \textit{Kepler} giant stars are known to harbor planets
discovered by detecting transits in the \textit{Kepler} light curves. 
\citet{Huber13} found two planets in the Kepler-56 system, while
\citet{Lillo-Box14} confirmed the hot-Jupiter Kepler-91~b. 
\citet{Ciceri15}, \citet{Ortiz15} and \citet{Quinn15} 
discovered the warm Jupiter Kepler-432~b. 
An additional planetary candidate has also been reported to Kepler-432
\citep{Quinn15} and Kepler-56 \citep{Otor16} found via a long-term RV
monitoring.
Here, we report on the discovery of a giant planetary candidate orbiting the K-giant 
star HD~175370 (KIC~007940959) in the \textit{Kepler} field found via the RV method.

\section{Observations and data analysis}\label{Observations}

We have observed HD~175370 since March 2010 using the coud\'e echelle
spectrograph at the 2-m Alfred Jensch Telescope of the TLS, Germany. 
We obtained 28 spectra with a S/N of
$\sim$ 100 per pixel in the extracted spectrum. 
Since May 2011 we have monitored this star using the fibre-fed
High Efficiency and Resolution Mercator Echelle Spectrograph (HERMES) 
at the 1.2-m Mercator Telescope on La Palma, Canary Islands, Spain. We
obtained
23 spectra with a S/N of $\sim$ 85 per pixel in the extracted spectrum. 
The RV measurements from both sites are listed in Table~\ref{table1}.

\begin{table}
  \begin{center}   
  \caption{RV and bisector velocity span (BVS) measurements of HD~175370
(see section \ref{activity2} for more details). 
The RVs were corrected for the barycenter motion. The RV uncertainty shown
here is an instrumental error.} \label{table1}
\begin{tabular}{lrrrr}
\hline
BJD & RV & $\sigma_{\rm{RV}}$ & BVS & $\sigma_{\rm{BVS}}$\\
(d) & ($\rm{m\,s^{-1}}$) & ($\rm{m\,s^{-1}}$) & ($\rm{m\,s^{-1}}$) & ($\rm{m\,s^{-1}}$)\\ 
\hline
\multicolumn{5}{l}{TLS RVs}\\
\hline
2455278.532376 &  -605.8 &  35.2 & & \\ 
2455278.602766 &  -542.6 &   8.3 & & \\ 
2455353.365205 &  -471.7 &   8.6 & & \\ 
2455451.296195 &  -328.7 &   8.0 & & \\ 
2455483.272526 &  -346.7 &   8.7 & & \\ 
2455488.271606 &  -353.9 &   7.6 & & \\ 
2455494.400977 &  -365.2 &   6.8 & & \\ 
2455666.577048 &  -353.7 &   8.7 & & \\ 
2455670.502131 &  -344.5 &   8.6 & & \\ 
2455677.567170 &  -325.1 &  10.1 & & \\ 
2455761.388294 &  -158.0 &   8.5 & & \\ 
2455849.373157 &  -219.0 &   7.9 & & \\ 
2456046.423323 &   -54.4 &  14.0 & & \\ 
2456051.588440 &    -9.1 &  11.1 & & \\ 
2456057.491668 &    -4.4 &  11.3 & & \\ 
2456090.435786 &    70.9 &   9.3 & & \\ 
2456096.490665 &    95.0 &   8.0 & & \\ 
2456101.446685 &   139.5 &  10.0 & & \\ 
2456167.316765 &     8.5 &  10.0 & & \\ 
2456169.423236 &    47.4 &   7.8 & & \\ 
2456229.271206 &  -163.6 &  11.5 & & \\ 
2456253.219143 &  -169.0 &   8.2 & & \\ 
2456407.417441 &   202.3 &  10.2 & & \\ 
2456412.527780 &   202.1 &   7.9 & & \\ 
2456458.380429 &   217.1 &   8.3 & & \\ 
2456461.561057 &   256.1 &   7.7 & & \\ 
2456496.364666 &   309.4 &   9.1 & & \\ 
2456549.517513 &   187.0 &   7.2 & & \\ 
\hline
\multicolumn{5}{l}{HERMES RVs}\\
\hline
2455685.659388 &  -13286 &     5 &  86 & 7 \\
2455770.401373 &  -13218 &     6 & 111 & 6 \\
2455790.522891 &  -13185 &     6 & 103 & 8 \\
2456039.673779 &  -13100 &     5 &  80 & 8 \\
2456125.521027 &  -12964 &     6 & 105 & 8 \\
2456160.424115 &  -13010 &     6 & 109 & 7 \\
2456397.618813 &  -12816 &     7 & 124 & 7 \\
2456457.504014 &  -12685 &     6 & 112 & 7 \\
2456555.460851 &  -12765 &     9 & 105 & 7 \\
2456564.427449 &  -12758 &    11 & 132 & 6 \\
2456565.448662 &  -12786 &     8 &  67 & 5 \\
2456598.291663 &  -12929 &    10 & 139 & 7 \\
2456604.286132 &  -12882 &     8 & 106 & 9 \\
2456607.290885 &  -12966 &    10 & 194 & 9 \\
2456754.705837 &  -12684 &     7 &  82 & 7 \\
2456804.590638 &  -12657 &    10 & 184 & 6 \\
2456846.419078 &  -12705 &     7 &  97 & 7 \\
2456884.486310 &  -12584 &    10 &  67 & 6 \\
2457212.518734 &  -12497 &    12 & 233 & 7 \\
2457214.736763 &  -12532 &     9 &  88 & 11\\
2457219.738180 &  -12534 &     7 & 119 & 10\\
2457220.734581 &  -12513 &     9 & 149 & 9 \\
2457222.643519 &  -12487 &     8 & 153 & 8 \\
\hline
  \end{tabular}
 \end{center}
\end{table} 

The coud\'e echelle spectrograph at TLS provides 
a wavelength range of 4670 -- 7400~\AA~and a spectral resolution of
67,000. We have reduced the data using standard \textsc{IRAF}\footnote{The Image
Reduction and Analysis Facility (\textsc{IRAF}) is 
distributed by the National Optical Astronomy Observatories, which are
operated by the Association of Universities for Research in Astronomy, Inc., under
cooperative agreement with the National Science Foundation.} procedures
(bias subtraction, flat-field correction, extraction of individual echelle
orders, wavelength calibration, subtraction of scattered light, cosmic rays removal and
spectrum normalization). The wavelength reference for the RV 
measurements was provided by an iodine absorption cell placed in the optical path just
before the slit of the spectrograph. The calculation of the
RVs largely followed the method outlined in \citet{Valenti95},
\citet{Butler96} and \citet{Endl00}, and takes into account changes in the instrumental
profile. We note that the measured RVs are 
relative to a stellar template which is an iodine-free spectrum 
(see Table~\ref{table1}) and are not absolute values.

For the HERMES spectrograph, we used a simultaneous ThArNe wavelength reference mode and 
low-resolution fibre (LRF) in order to achieve as accurate RV 
measurements as possible. The LRF mode has a spectral resolution
of 62,000 and HERMES has a wavelength range of 3770 -- 9000~\AA. More details about 
the HERMES spectrograph can be found in
\citet{Raskin11}. We have used a dedicated automated data reduction pipeline
and RV toolkit (\textsc{HermesDRS}) to reduce the data and calculate
absolute RVs (see Table~\ref{table1}). The spectral mask of Arcturus on 
the velocity scale of the IAU RV standards was used for the cross-correlation.

The two data sets (relative TLS and absolute HERMES RVs) were combined adjusting
their zero points during the orbital fitting procedure as described in section
\ref{orbit}.  

\section{Properties of the star HD~175370}\label{properties}

HD~175370 (KIC~007940959) has a visual magnitude of $m_V=7.19$ mag
\citep{Hog00}. The parallax was determined by \citet{Leeuwen07} from
\textit{Hipparcos} data as $3.68\pm 0.45$ mas which implies an absolute magnitude 
$M_V=0.02\pm 0.27$ mag. Table~\ref{table2} lists the stellar parameters of
HD~175370 known from literature together with those determined in this work. 

\begin{table}\tabcolsep=3.5pt
  \begin{center}   
  \caption{Stellar parameters of HD~175370. The top part of the table lists
parameters that are derived using spectral analysis or calculated in this
paper, unless stated otherwise. The central part of the table shows resulting 
parameters from asteroseismic analysis. The bottom part of
the table lists derived parameters using \textsc{GBM} modelling as described in
section \ref{properties2}. We show parameters for the red-giant branch
(RGB), red clump (RC) and asymptotic-giant branch (AGB) star.
The probability for each evolutionary phase gives the total 
integrated likelihood of the distributions.} \label{table2}
\begin{tabular}{lcccc}
\hline
Parameter & \multicolumn{3}{c}{Value} & Unit\\
\hline
$m_V^a$ & \multicolumn{3}{c}{$7.19\pm 0.01$} & mag \\
$B-V^a$ & \multicolumn{3}{c}{$1.27\pm 0.02$} & mag \\
Parallax$^b$ & \multicolumn{3}{c}{$3.68\pm 0.45$} & mas \\
$M_V$ & \multicolumn{3}{c}{$0.02\pm 0.27$} & mag \\
Distance & \multicolumn{3}{c}{$272\pm 33$} & pc \\
Spectral type & \multicolumn{3}{c}{K2$\,$III} & \\
$T_{\rm{eff}}$ & \multicolumn{3}{c}{$4301\pm 43$} & K\\
$[\rm{Fe/H}]$ & \multicolumn{3}{c}{$-0.52\pm 0.07$} & dex\\
$v_{\rm{turb}}$ & \multicolumn{3}{c}{$1.63\pm 0.13$} & $\rm{km\,s^{-1}}$\\
$v\,\sin i$ & \multicolumn{3}{c}{$6.11\pm 0.50$} & $\rm{km\,s^{-1}}$\\
$\log g$ & \multicolumn{3}{c}{$1.70\pm 0.15$} & dex\\
\hline 
$\Delta\nu$ & \multicolumn{3}{c}{$6.7\pm 0.2$} & $\mu$Hz\\ 
$\nu_{\rm max}$ & \multicolumn{3}{c}{$1.17\pm 0.03$} & $\mu$Hz\\
\hline
& RGB star & RC star & AGB star & \\
\hline
$\log g$   & $1.70\pm 0.02$      & $1.700\pm 0.004$     & $1.71\pm 0.02$      & dex\\
$M_{\ast}$ & $1.02\pm 0.16$      & $0.85\pm 0.04$       & $1.04\pm 0.15$      & $M_{\odot}$\\
$R_{\ast}$ & $23.5\pm 3.4$       & $21.6\pm 0.4$        & $24.0\pm 3.2$       & $R_{\odot}$\\
Age        & $8.7^{+5.0}_{-6.0}$ & $11.5^{+1.6}_{-1.4}$ & $5.4^{+7.9}_{-3.2}$ & Gyr\\
$L_{\ast}$ & $173^{+32}_{-27}$   & $155\pm 2$           & $194^{+29}_{-25}$   & $L_{\odot}$\\
Probability & 5551.3 & 0.9 & 277.0 & \\
\hline
  \end{tabular}

 \medskip
$^a$\citet{Hog00} $^b$\citet{Leeuwen07}
 \end{center}
\end{table} 

\subsection{Spectral analysis}\label{properties1}

The basic stellar parameters were determined from two high-resolution
spectra of HD~175370. One was taken with the high-resolution fibre (HRF) mode of the 
HERMES spectrograph (R=85,000) with a S/N of 150. Another spectrum was
obtained with the TLS spectrograph without the iodine cell (R=67,000) with a
S/N of 150. 

For the analysis we used the spectrum-synthesis method, comparing the observed
spectrum with a library of synthetic spectra computed from atmosphere models 
on a grid of stellar parameters. We used \textsc{SynthV} \citep{Tsymbal96} for computing
the synthetic spectra. \textsc{SynthV} is a spectrum synthesis code based on plane-parallel 
atmospheres and working in a non-local thermodynamic equilibrium (NLTE) regime. 
It has the advantage that for each chemical 
element different abundances can be considered. 

The free stellar parameters in our analysis were the stellar effective
temperature, $T_{\rm{eff}}$, the stellar gravity, $\log g$, the
solar-scaled abundance, $[\rm{M/H}]$, the microturbulent velocity, 
$v_{\rm{turb}}$, and the
projected rotational velocity, $v\,\sin i$. The goodness of
the fit as well as the parameter uncertainties were calculated from
$\chi^2$-statistics \citep{Lehmann11}.

In particular in the blue spectral region it is difficult to do a proper 
normalization of the observed spectra to the true local continuum of such 
a late-type star. Although our spectral analysis includes an adjustment 
of the continuum of the observed spectrum to that of the synthetic one, 
we observed that larger uncertainties in the parameters arise from extending
the region that is used too far into the blue wavelengths. Finally, we restricted the 
spectral range to 4550 -- 6860~\AA~for the HERMES spectrum. For the TLS spectrum, 
we used the wavelength region 4705 -- 6860~\AA. In both cases we
excluded some small regions where strong telluric lines occur.

We tested different atmosphere models like Kurucz plane-parallel models
\citep{Kurucz93} and MARCS spherical models \citep{Gustafsson08}
that were interpolated onto a finer parameter grid. We
decided to use standard composition MARCS models for our final analysis.

The analysis was performed in several steps. First, we optimized
the free parameters as mentioned before. Then we optimized the iron abundance 
together with the microturbulent velocity. In the next step, we
optimized the abundances of individual elements and finally all
steps were repeated in an iterative way.

Table~\ref{table2} lists the stellar parameters that we derived. Because of the
degeneracy between the single parameters, their optimum values
and the uncertainties have been calculated from the $\chi^2$-hypersurface including 
all grid points in all parameters. 

The TLS spectrum delivered $\log g=1.70$ and the HERMES 
spectrum $\log g=1.25$, whereas the values of all other parameters like 
$T_{\rm{eff}}$, $[\rm{Fe/H}]$, $v_{\rm{turb}}$, and $v\,\sin i$ agreed within the 
1-$\sigma$ error bars. The value of $\log g$ derived from the TLS spectrum 
agrees very well with the value obtained from our asteroseismic analysis
(see Table~\ref{table2}). 
Restricting the wavelength range for the HERMES spectrum to the same range 
as used for the TLS spectrum gave $\log g=1.5$. We conclude that the determination 
of $\log g$ is very sensitive to the quality of spectrum normalisation and
assume that the continuum of the pipeline-reduced HERMES spectrum is the more uncertain 
the more blue wavelengths are considered. The derived $T_{\rm{eff}}$ and $\log g$ 
correspond to a K2$\,$III star.

\begin{table}\tabcolsep=3pt
  \begin{center}   
  \caption{Abundances of HD~175370 relative to the solar composition together with
the lower and upper error bars.} \label{table3}
\begin{tabular}{ccccc}
\hline
C & O & Na & Mg & Ca\\
$-0.49^{+0.14}_{-0.09}$ & $-0.55^{+0.09}_{-0.28}$ & $-0.34^{+0.26}_{-0.23}$ &
$-0.62^{+0.08}_{-0.08}$ & $-0.56^{+0.17}_{-0.16}$\\
\hline
Ti & V & Cr & Fe & Ni\\
$-0.44^{+0.12}_{-0.11}$ & $-0.52^{+0.18}_{-0.17}$ & $-0.55^{+0.17}_{-0.16}$ &
$-0.52^{+0.07}_{-0.07}$ & $-0.48^{+0.17}_{-0.16}$\\
\hline
  \end{tabular}
 \end{center}
\end{table} 

Finally, we used the TLS spectrum to determine the abundances of
all chemical elements that show significant line contributions in
the wavelength range that we considered. 
The results are listed in Table~\ref{table3}.

The investigation of a large sample of giant stars in the local region by
\citet{Luck07} shows that there is a general trend of $[\rm{C/Fe}]$
to reach zero value for low-metallicity stars around $[\rm{Fe/H}]=-0.5$, 
accompanied by an increase of $[\rm{O/Fe}]$ to about 0.4 (unfortunately, no indication of
stellar masses was given by the authors for these trends). 
We could determine the carbon and oxygen abundances from our spectra 
and thus check if our results fit into this general trend. Both the
C and O abundances correspond to the derived Fe abundance, 
i.e. $[\rm{C/Fe}]$ and $[\rm{O/Fe}]$ are about zero. Since spherical 
MARCS atmosphere models are available as standard and CN-cycled models, 
we additionally used CN-cycled models to check for possible effects. 
The calculations showed, however, that the influence of using such different 
models on the results is marginal. 

\subsection{Asteroseismic analysis}\label{properties2}

HD~175370 was observed throughout the entire \textit{Kepler} mission in
long-cadence mode with a 29.4-minute sampling rate. We used the raw data to
which corrections for instrumental effects have been applied
in the same manner as in \citet{Garcia2011}. The concatenated data set was then
filtered with a triangular filter with a full-width at half maximum of
24 days to remove any remaining instrumental effects.

The \textsc{Octave} (Birmingham - Sheffield Hallam) automated pipeline \citep{hekker2010} 
was then used to determine the large
frequency separation between modes of consecutive order and same degree
($\Delta\nu$) and the frequency of maximum oscillation power ($\nu_{\rm max}$).
We obtained 6.7 $\pm$ 0.2 $\mu$Hz and 1.17 $\pm$ 0.03 $\mu$Hz for
$\Delta\nu$ and $\nu_{\rm max}$, respectively. These values are combined in
a grid-based modelling (\textsc{GBM}) effort using $T_{\rm eff}$ and [Fe/H] from the
spectroscopic measurements (see Table~\ref{table2}) to determine stellar mass,
radius, and age. We used a \textsc{GBM} code \citep{hekker2013}
with the canonical Bag of Stellar Tracks and Isochrones (BaSTI) models 
\citep{pietrinferni2004} and a maximum likelihood
estimation based on \citet{basu2010}. 
As this star has sub-solar metallicity we accounted for that by using the metallicity
dependent reference value of \citet{Guggenberger16} to compute
$\Delta\nu$ from the models via the scaling relation mentioned in section
\ref{intro}. In the \textsc{GBM} code we have implemented
the possibility to model stars on the red-giant branch (RGB,
hydrogen-shell burning phase), in the red clump (RC, hydrogen-shell and
helium-core burning), and on the asymptotic-giant branch (AGB, hydrogen-shell
and helium-shell burning) separately. 
Using this option we find results for all three evolutionary stages (see
Table~\ref{table2}). Nevertheless, the probability distributions indicate 
that this star is not a RC star, but most likely it is a RGB or AGB star. 
Henceforth, we only provide these two solutions throughout the paper.
Also, according to \citet{Mosser14}, HD~175370 is not a RC
star, and most likely it is a RGB star.

\section{Planetary orbital solution}\label{orbit}

\begin{figure}
\includegraphics[width=88mm]{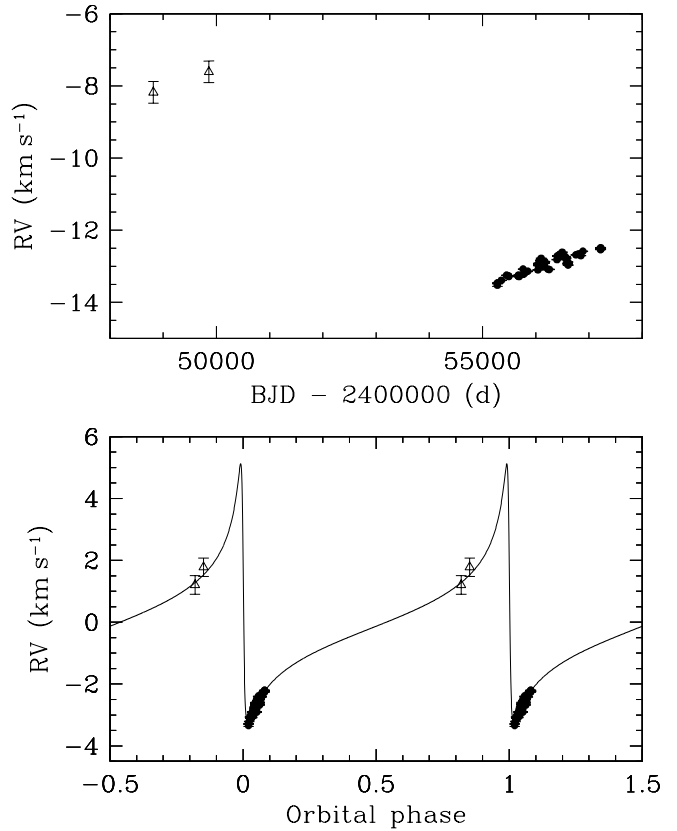}
\caption{Top: The RV measurements of HD~175370 and their error bars from
this work (filled circles) and from \citet{Famaey05} (open triangles). 
Bottom: The RV measurements of HD~175370 and their error bars as above phased to the
orbital period of 32,120 d with an orbital solution overplotted (solid curve).}
\label{plot1}
\end{figure}

\begin{figure}
\rotatebox{270}{\includegraphics[width=66mm]{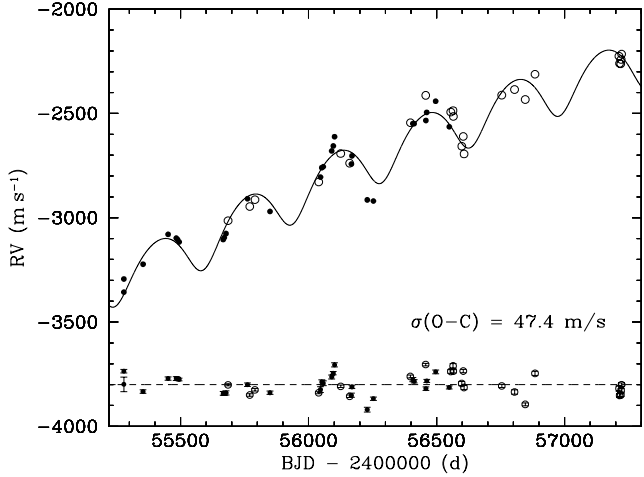}}
\caption{The RV measurements of HD~175370 obtained from March 2010 to July 2015 
using the coud\'e echelle spectrograph at
TLS, Germany (filled circles) and the HERMES spectrograph at Mercator, La Palma (open circles).
The solid curve is the superposition of the binary and planetary orbital
solutions (see Table~\ref{table4}). The RV residuals and the error bars 
after removing the binary and planetary orbital solutions are shown in the lower
part.}
\label{plot2}
\end{figure}

We monitored HD~175370 for a time span of 5 years and 4 months, during which we acquired
51 RV measurements (see Table~\ref{table1} and filled circles in
Fig.~\ref{plot1}). 
\citet{Famaey05} reported a RV of HD~175370 of $-7.90\pm 0.28\,\,\rm{km\,s^{-1}}$,
and we show their measurements in Fig.~\ref{plot1} as open triangles. Our RV
measurements show a long-term trend and also changes which could be caused by a planetary companion.

In order to find a planetary orbital solution we first fitted a binary orbit which was then 
subtracted from the data. We used the non-linear least squares fitting program 
\textsc{Gaussfit} \citep{McArthur94}. For the
binary orbit, due to the high eccentricity and poor data sampling, convergence 
did not occur when we let the orbital period, $P$, and eccentricity, $e$, as free
parameters. Instead, we fixed the period and eccentricity of the binary
to values obtained from varying these parameters separately and searched for a
minimum in the residuals. We fitted RV zero points of the three data sets
(Mercator, TLS, and \citet{Famaey05}), periastron epoch, $T_{\jed{periastron}}$, 
periastron longitude, $\omega$, and semi-amplitude of the RV curve, $K$. 

Originally, to find a planetary orbital solution, we fitted a linear fit to
our RV measurements instead of the binary orbit. When we collected more RVs
it became clear that the trend of our RV data is not linear, and that a binary
orbital solution is a better approximation despite the fact that errors on
the orbital period and eccentricity must be large.

Due to the higher uncertainties of the RV measurements of \citet{Famaey05},
we used only Mercator and TLS data to obtain the planetary orbital solution. 
We fitted simultaneously the orbital period, periastron epoch, 
eccentricity, periastron longitude, semi-amplitude of the RV curve, and RV
zero points of the two data sets from TLS and Mercator. The latter was
possible because parts of the data have been taken close in time.

\begin{figure}
\rotatebox{270}{\includegraphics[width=66mm]{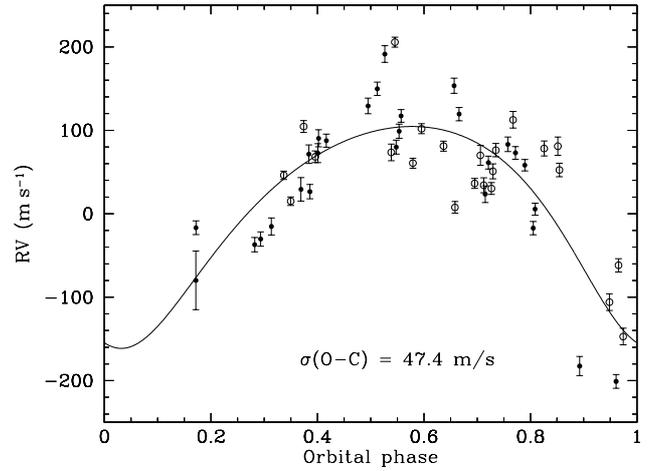}}
\caption{The RV measurements of HD~175370 after removing the binary orbit 
plotted with corresponding error
bars phased to the orbital period of 349.5 d with an orbital solution overplotted
(solid curve). Filled and open circles are for RVs from TLS and
Mercator, respectively.}
\label{plot3}
\end{figure}

We then subtracted the planetary orbital solution from the original RVs and made
another binary orbital solution, which resulted in lower uncertainties of
all fitted parameters. We used this binary orbital solution to find a
new planetary orbital solution, and we repeated the above procedure one more time.  
The results after iterative pre-whitening of the RVs for the binary solution and 
the planet solution and vice versa are shown in Table~\ref{table4}. 
The phase-folded RV variations for the binary orbital solution and the orbital
fit are shown in Fig.~\ref{plot1} (bottom panel). 
In Fig.~\ref{plot2} we show the RV measurements from
TLS and Mercator with the superposition of the planetary and binary
orbital solutions. Fig.~\ref{plot2} (lower part) also shows the RV residuals
after removing the binary and planetary orbits. A periodogram analysis of the residual RV 
data showed no additional significant frequencies. Finally, the phase-folded RV variations 
for the planetary orbital solution and the orbital fit are shown in
Fig.~\ref{plot3}. 

In order to take into account intrinsic stellar jitter we
used a scatter of RV residuals instead of instrumental errors as RV
uncertainties. This resulted in an increased uncertainty for the orbital
period and semi-amplitude of the RV curve, but we believe that it is a more realistic estimate
of real uncertainties. 
The scatter of RV residuals is also similar to the expected
jitter based on stellar properties (see below), and therefore using it as RV uncertainties
is more justifiable.
To check the uncertainty in the orbital period, we
have used a formula of \citet{Horne86} for estimating the uncertainty in the
period from a peak in the periodogram, and got an uncertainty of 2.7~d.
We also took the orbital solution sampled in the same way as the data and
added random noise at $\sigma=50\,\,\rm{m\,s^{-1}}$. We then used different
seeds and the rms scatter of the final periods was 4.3 d. Our orbital period 
uncertainty from the \textsc{Gaussfit} fitting procedure, 4.5 d, is consistent with
the other two estimates, and thus we believe is realistic.

Our planetary orbital solution yields a mass function of
$f(m)=(0.79\pm 0.48)\times 10^{-7}\,M_{\odot}$. 
Using our derived stellar
mass for a RGB and AGB star (see Table~\ref{table2}), the minimum mass
for the companion is $M\,\sin i=4.6\pm 1.0\,M_{\rm{Jupiter}}$. 

\begin{figure}
\rotatebox{270}{\includegraphics[width=66mm]{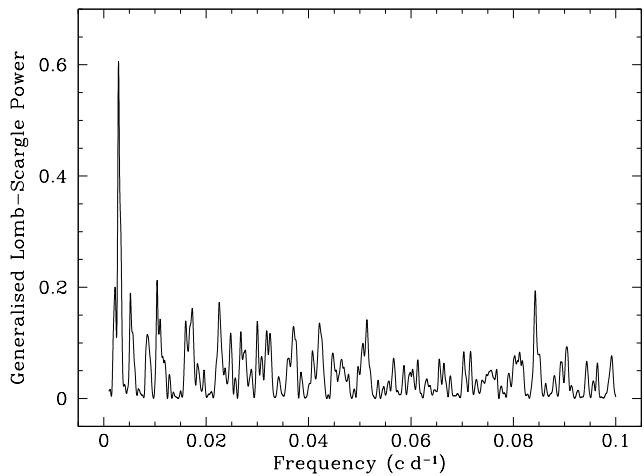}}
\caption{A generalised Lomb-Scargle periodogram of the HD~175370 RV
measurements after removing the binary orbit. There is a very high peak 
with a Scargle power of 15.14 at a
frequency $\nu=0.002881\jed{c\,d^{-1}}$, corresponding to
a period of $P=347.1\jed{d}$.}
\label{plot4}
\end{figure}

A generalised Lomb-Scargle (GLS) periodogram \citep{Zechmeister09} of the RV
measurements after removing the binary orbital solution is shown in
Fig.~\ref{plot4}. 
There is a statistically significant peak at a frequency of $\nu=0.002881\jed{c\,d^{-1}}$, 
corresponding to a period of $P=347.1\jed{d}$. The peak has a normalized power of 0.606 which
translates to a Scargle power \citep{Scargle82} of 15.14. The false alarm probability (FAP) 
of the 347.1-d period was estimated using a bootstrap randomization technique 
\citep{Kurster97}. For that, the RV values were
randomly shuffled keeping the times fixed and the GLS periodogram 
was calculated for each random data set over the frequency
range as shown in Fig.~\ref{plot4}. The fraction of
random periodograms that had a Scargle power greater than the
data periodogram gave us an estimate of the FAP that the signal is due to noise. 
After 100,000 shuffles there was no random periodogram having a power larger than the real data
set, which confirms that the periodic signal is not due to noise or data
sampling. In addition, the derived planetary orbital period of
$P=349.5\pm 4.5\jed{d}$ agrees
very well with the period found from the GLS periodogram. 

We note that the scatter of the RV residuals after fitting the binary and planetary
orbits is 47.4$\,\,\rm{m\,s^{-1}}$, significantly higher than
instrumental errors of our RV measurements. This scatter arises from stellar oscillations. According
to the scaling relations of \citet{Kjeldsen95}, the velocity amplitude for stellar 
oscillations is expected to be 
$v_{osc}=((L/L_{\odot})/(M/M_{\odot}))\,23.4\,\rm{cm\,s^{-1}}$. 
\citet{McDonald12} obtained the stellar luminosity of
$L_{\ast}=143.75\,L_{\odot}$. They also derived the stellar
effective temperature of $T_{\rm{eff}}=4424\,\rm{K}$. 
We derived the stellar luminosity and the stellar mass using 
\textsc{GBM} modelling (see Table~\ref{table2}), which results in 
a velocity amplitude of $v_{osc}=39.7\pm 9.6\,\rm{m\,s^{-1}}$ for a RGB star
and $v_{osc}=43.7\pm 9.1\,\rm{m\,s^{-1}}$ for an AGB star. 
In both cases the resulting
velocity amplitude is consistent with our RV rms scatter within the
1-$\sigma$ error bars.

Also, a large scatter of the RVs is expected for HD~175370 due to its
$B-V$ color of 1.27 mag. \citet{Hekker06} studied 179 K giants and confirmed the increase of 
the RV variability with $B-V$ color, first described by \citet{Frink01}.

\begin{table}\tabcolsep=3pt
  \begin{center}   
  \caption{Orbital parameters of HD~175370~B and HD~175370~b.} \label{table4}
\begin{tabular}{lccccc}\hline
Parameter & \multicolumn{4}{c}{Value} & Unit\\
\hline
          & \multicolumn{2}{c}{HD~175370~B} & \multicolumn{2}{c}{HD~175370~b} & \\
\hline
Period & \multicolumn{2}{c}{32120 (fixed)} &
\multicolumn{2}{c}{$349.5\pm 4.5$} & d\\
$T_{\jed{periastron}}$ & \multicolumn{2}{c}{$2454614\pm 100$} &
\multicolumn{2}{c}{$2456267\pm 30$} & BJD\\
$K$ & \multicolumn{2}{c}{$4181\pm 99$} &
\multicolumn{2}{c}{$133\pm 25$} & $\jed{m\,s^{-1}}$\\
$e$ & \multicolumn{2}{c}{0.879 (fixed)} &
\multicolumn{2}{c}{$0.22\pm 0.10$} & \\
$\omega$ & \multicolumn{2}{c}{$75\pm 1$} &
\multicolumn{2}{c}{$162\pm 28$} & deg\\
$\sigma$(O-C) & \multicolumn{2}{c}{} &
\multicolumn{2}{c}{47.4} & $\jed{m\,s^{-1}}$\\ 
$f(m)$ & \multicolumn{2}{c}{0.026} &
\multicolumn{2}{c}{$(0.79\pm 0.48)\times 10^{-7}$} & $M_{\odot}$\\
\hline 
RV$_0^{\rm{Famaey}}$ & \multicolumn{2}{c}{-9372} & \multicolumn{2}{c}{} & $\jed{m\,s^{-1}}$\\
RV$_0^{\rm{TLS}}$ & \multicolumn{2}{c}{-10167} &
\multicolumn{2}{c}{-60} & $\jed{m\,s^{-1}}$\\
RV$_0^{\rm{HERMES}}$ & \multicolumn{2}{c}{-10272} &
\multicolumn{2}{c}{-55} & $\jed{m\,s^{-1}}$\\
\hline
& RGB$^a$ & AGB$^b$ & RGB$^a$ & AGB$^b$ & \\
\hline
$a$ & 22 & 22 & $0.98\pm 0.05$ & $0.98\pm 0.05$ & AU \\
$M\,\sin i$ & & & $4.6\pm 1.0$ & $4.6\pm 1.0$ & $M_{\rm{Jupiter}}$\\
$M\,\sin i$ & 0.37 & 0.38 & & & $M_{\odot}$\\
\hline
  \end{tabular}

 \medskip
Assuming HD~175370 is a $^a$red-giant branch or $^b$asymptotic-giant branch
star.
 \end{center}
\end{table} 

\section{Stellar activity analysis}\label{activity}

RV searches for planets are affected by intrinsic stellar variability
causing RV changes which are not related to a stellar reflex motion due to an orbiting
planet. \citet{Hatzes93} showed that the low-amplitude, long-period
RV variations may be attributed to pulsations, stellar activity or low-mass
companions, while the presence of low-amplitude, short-period RV variations
are due to p-mode oscillations \citep{Hatzes94}. 
\citet{Hekker08} concluded that intrinsic mechanisms play an important role
in producing RV variations in K-giant stars, as suggested by their dependence on
$\log g$, and that periodic RV variations are additional to these intrinsic variations, 
consistent with them being caused by companions.

Therefore, in order to confirm the existence of a planet, it is important 
to investigate the origin of the RV variations in detail. 
The solar-type p-mode oscillations observed in the high-quality \textit{Kepler}
photometry have all much shorter periods than 349.5 days and can be ruled out as an explanation 
for the observed long-period RV variations of HD~175370. We have investigated \textit{Kepler} light
curves and measured spectral line bisectors and chromospheric activity to check whether stellar activity
could be a reason for the observed RV changes.

\subsection{Photometric variations}\label{activity1}

\begin{figure}
\includegraphics[width=88mm]{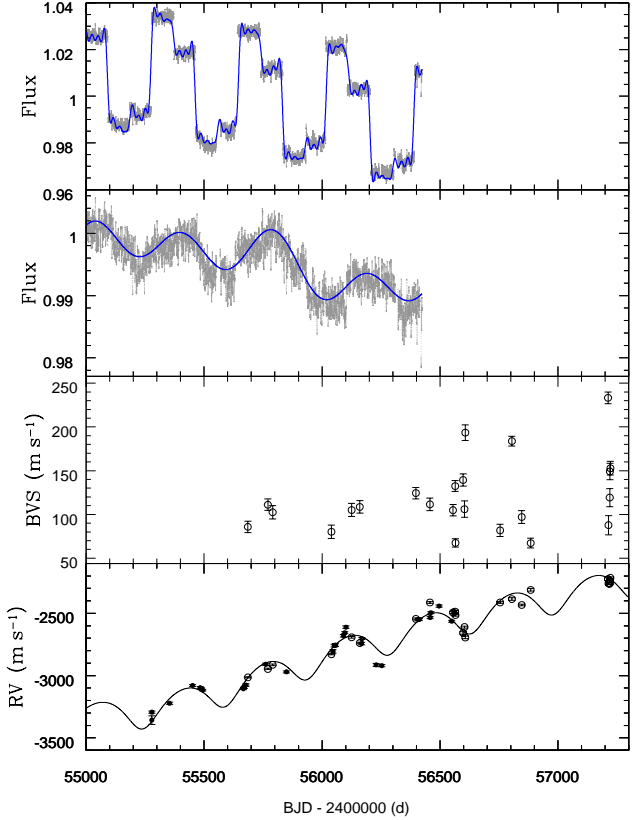}
\caption{Top: The raw \textit{Kepler} data of quarters Q2 -- Q17 (grey
curve). The blue curve is the sum of sine functions of periods found by
\textsc{Period04}.
Second from the top: The merged \textit{Kepler} light curve of quarters Q2 -- Q17
(grey curve) where one
can see both long-period changes and short-period changes due to
stellar oscillations. The blue curve is the sum of sine functions of three
long periods found by \textsc{Period04}. Second from the bottom: 
Bisector velocity spans (BVS) measured from the average bisector values
between flux levels of 0.1 -- 0.3 and 0.7 -- 0.9 of the continuum value, 
shown with corresponding error bars. Bottom: The RV variations of HD~175370 
(TLS: filled circles, HERMES: open circles) shown for comparison.}
\label{plot5}
\end{figure}

\begin{figure}
\includegraphics[width=88mm]{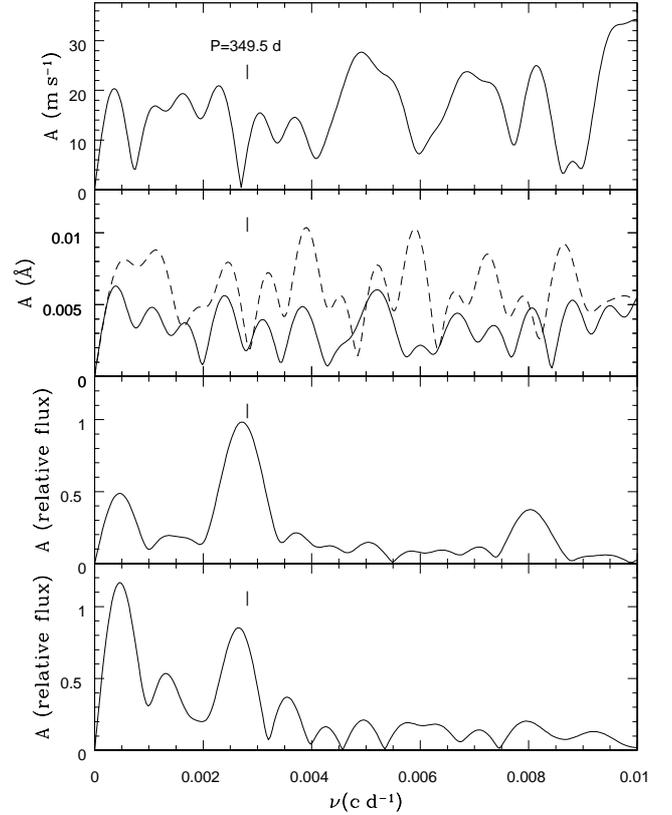}
\caption{Top: The frequency spectrum of bisector velocity spans. Second from
the top: The frequency spectrum of \ion{Ca}{ii} 8662 \AA~spectral line 
equivalent width measurements (solid line is the frequency spectrum for 
equivalent width measurements from 8661.0 to 8663.27 \AA~and dashed line 
from 8661.55 to 8662.8 \AA). Second from the bottom: The frequency spectrum
of the \textit{Kepler} raw data without aligning individual quarters.
Bottom: The frequency spectrum of merged \textit{Kepler} data. 
See text for more details. A is the amplitude of the frequency spectra.
The vertical line corresponds to the orbital period of the planetary
candidate.} 
\label{plot6}
\end{figure}

To check whether rotational modulation might be a cause of the observed
RV variations, we analysed the \textit{Kepler} photometry of HD~175370.
The star was observed by the \textit{Kepler} satellite \citep{Koch10} since May 2009 and for all
quarters during the main \textit{Kepler} mission. We used data
corrected by the \textit{Kepler} science team and merged individual quarters
using a constant fit to the last and first two days of each quarter. The second
quarter, Q2, was our reference. The resulting light curve is shown in the
second panel from the top of Fig.~\ref{plot5}. We
searched for periods using the program \textsc{Period04} \citep{Lenz05},
where multiple periods can be found via a pre-whitening procedure. A Fourier
analysis is used to identify the dominant period in the data which is then fitted by a
sine wave. Each found period is then subtracted, and an additional search
for periods is made in the residuals by a further Fourier analysis. We performed
a Fourier analysis repeatedly to extract significant peaks corresponding
to long-term periods. We found three long-term periods in the data:
$P_1=2186.5\jed{d}$, $P_2=389.4\jed{d}$, and $P_3=676.8\jed{d}$. The sum of sine
functions of individual periods is over-plotted in the
second panel from the top of Fig.~\ref{plot5} 
as a blue curve. The zoom of the frequency spectrum is shown in the
bottom panel of Fig.~\ref{plot6}, and one can see that there is indeed a peak at the
frequency of 0.00257 c/d, corresponding to a period of 389.4 d. 

The period of 389.4 d is close to the orbital period of the \textit{Kepler}
satellite which is 372.5~d. In order to access whether this period is
real, we have analysed raw light curves in the same way as above without
aligning individual quarters. The raw \textit{Kepler} data and the sum of sine
functions of periods found by \textsc{Period04} are shown in the top panel of
Fig.~\ref{plot5} and the frequency spectrum is displayed in the second panel from 
the bottom of Fig.~\ref{plot6}. We found a dominant period of 370 d which agrees 
very well with the orbital period of the \textit{Kepler} satellite. 

Every 3 months, the \textit{Kepler} satellite was reoriented (rotated one-quarter turn) 
and then the light from each target star was collected by a new set of pixels on a different 
CCD. For precision photometry, this will introduce systematic errors. 
In addition, the orbital period of the \textit{Kepler} satellite, 372.5~d, 
will also introduce systematic errors in the relative photometry. Therefore,
finding a period in the \textit{Kepler} data which is close to the orbital 
period of the satellite is suspicious. If one would
like to use \textit{Kepler} light curves to find periods of real physical
processes close to the orbital period of the satellite, one would have to look
at all K giants in the field and detrend them using a program like
\textsc{Sys-Rem} \citep{Mazeh07}. Such systematic detrending would be useful for HD~175370, however, it is beyond 
the scope of this paper. Instead, we have looked at \textit{Kepler} 
light curves of all 41 giant stars from our sample which we follow from Tautenburg 
and Mercator and which have corrected timeseries by the KASC Data Analysis Team. 76 per cent
of targets show the dominant period of $390\pm 20\jed{d}$, implying that the
orbital period of the \textit{Kepler} satellite can be really seen in the
relative photometry of most of the giant stars. 7 per cent show the
second dominant period of $379\pm 21\jed{d}$. Finally, 17 per cent of targets show
dominant periods which are clearly related to real physical processes and are not close to
the orbital period of the satellite.   

\begin{figure}
\rotatebox{270}{\includegraphics[width=66mm]{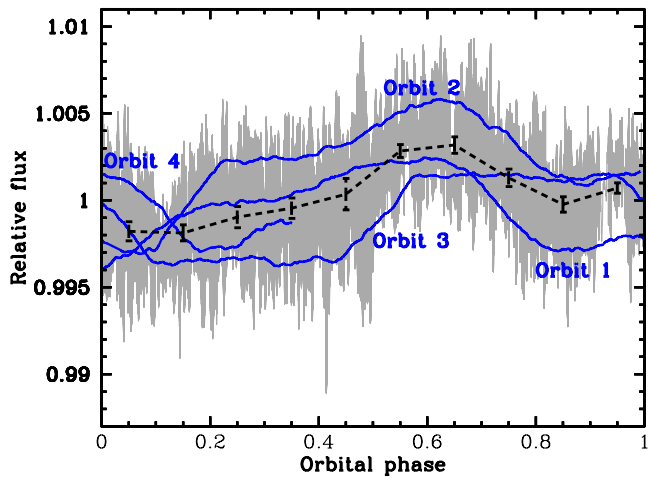}}
\caption{The \textit{Kepler} light curve phase-folded to the orbital period
of the planet (grey solid curve) and a running mean over 50 d (blue solid curve), where 
order of phase-folded orbits is shown. The black dashed curve shows the average 
values of the relative flux in bins of width 0.1 in phase. The error bars
along the black dashed curve correspond to the standard deviation in each phase bin.}
\label{plot10}
\end{figure}

\begin{figure}
\includegraphics[width=88mm]{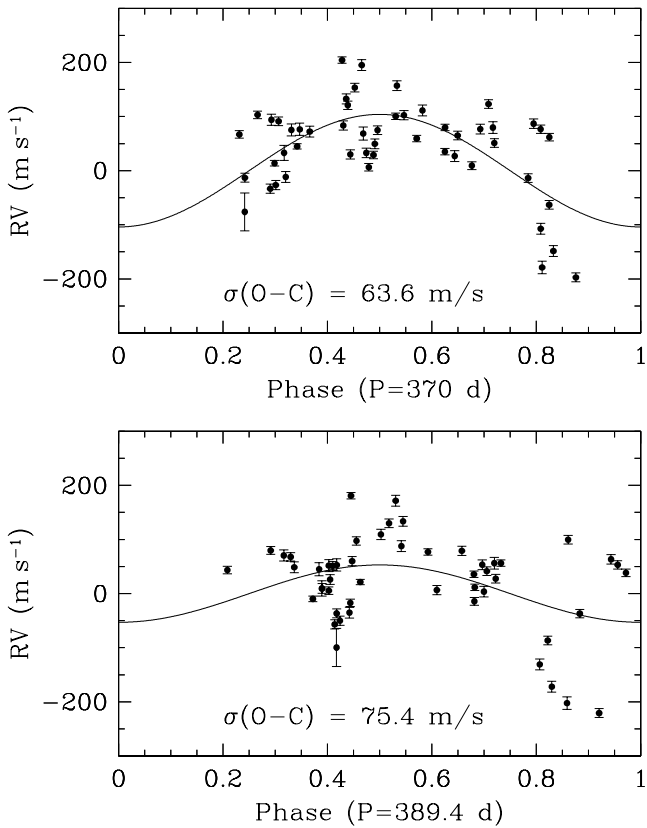}
\caption{The RV data phased to the photometric periods of 370~d and 389.4~d. 
The solid curve represents a best fit allowing periastron epoch and semi-amplitude 
of the RV curve to vary and assuming a constant period and zero
eccentricity.}
\label{plot11}
\end{figure}

Both periods of 389.4 d and 370 d are also similar to the planetary orbital
period of 349.5 d. We subtracted a linear trend from the
photometric data shown in the second panel from the top of Fig.~\ref{plot5}
and phase-folded them to the orbital period of the planet. The result is shown in
Fig.~\ref{plot10}, where the black dashed curve connects average values of the relative
flux in bins of width 0.1 in phase. There is a modulation with an amplitude of 
about 0.0021 mag. The photometric time series cover nearly 3.5
planetary orbits. A careful look reveals that during 
the first two planetary orbits when photometric
data were taken, photometric and RV variations agree quite well in phase, 
and only since the third orbit photometric and RV variations differ from each 
other (see Fig.~\ref{plot10}). We have also phased the RV data to both 
photometric periods of 389.4~d and 370~d, and results are shown in
Fig.~\ref{plot11}. The solid curve represents a best fit allowing periastron
epoch and semi-amplitude of the RV curve to vary and assuming a constant
period and zero eccentricity. It can be seen that the RV data do not phase
well with the photometric periods 370 and 389.4 d, having
the scatter of RV residuals 63.6 and 75.4$\,\,\rm{m\,s^{-1}}$,
respectively. Even though the above is not a proof that
the RV period is unrelated to the photometric periods, there is no evidence
to contradict this when taking into account that the \textit{Kepler} photometry
is suspicious for periods close to the orbital period of the satellite.

In addition, a stellar rotation period is expected
to be shorter than the period of the planetary candidate. With the measured  projected rotational velocity 
of $v\,\sin i=6.11\pm 0.50\,\rm{km\,s^{-1}}$
and the radius of $R_{\ast}=23.5\pm 3.4\,R_{\odot}$ (RGB star) and
$R_{\ast}=24.0\pm 3.2\,R_{\odot}$ (AGB star), the expected
upper limit for the stellar rotation period is 194$^{+48}_{-40}$ d and
199$^{+46}_{-40}$ d, respectively.
It is unlikely that the 349.5-d RV period represents the rotation period
of the star. Assuming it did, this would result in a {\it maximum} projected rotational
velocity of 3.4 $\pm$ 0.5 $\rm{km\,s^{-1}}$ (RGB star) and 3.5 $\pm$ 0.5 $\rm{km\,s^{-1}}$
(AGB star), or nearly a factor of two smaller than our measured $v\,\sin i$.

Furthermore, if the apparent sinusoidal photometric variations in the second
panel from the top in Fig.~\ref{plot5} 
are due to cool spots, they cannot explain the observed RV variation.
The amplitude of the photometric variations is
$\sim$~0.21 per cent. This results in a spot-induced RV amplitude of $\sim$~11
m\,s$^{-1}$ \citep{Saar97, Hatzes02}, or  more than a factor of
ten smaller than the observed $K$-amplitude. 

\subsection{Spectral line bisector analysis}\label{activity2}

\begin{figure}
\includegraphics[width=88mm]{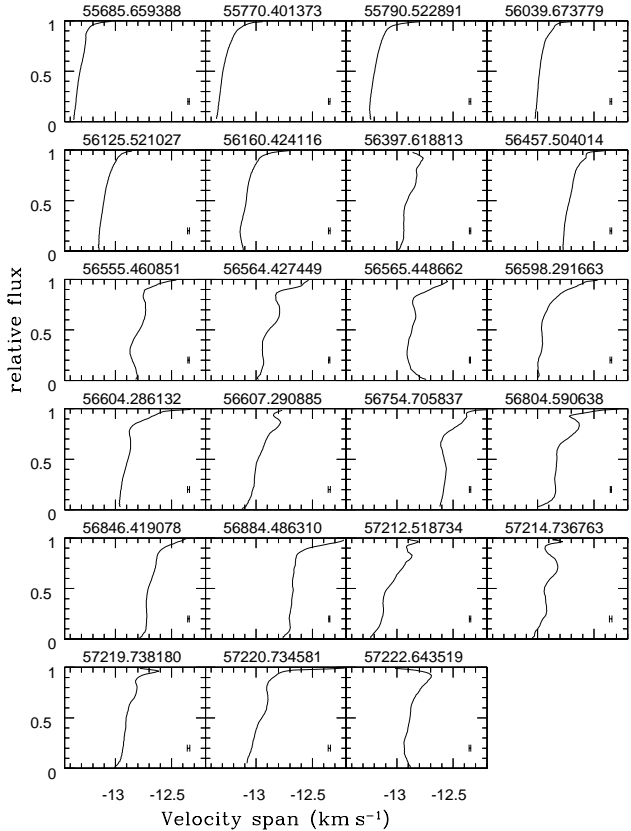}
\caption{Spectral line bisectors for all spectra of HD~175370 observed by
the HERMES spectrograph, measured from the cross-correlation function based on $\sim$1650
spectral lines. BJD of the observation is shown on top of each bisector plot.
Bisector velocity span errors are shown at lower right of each plot.}
\label{plot7}
\end{figure}

Stellar rotational modulations of inhomogeneous surface features can create
variable asymmetries in the spectral line profiles. The best way to describe
this asymmetry is with the line bisector, which consists of the midpoints 
of horizontal line segments extending across the line profile. In a typical case, due to 
stellar granulation, the red wing of the stellar line is depressed, which causes 
the bisector to have a positive slope and curve to the right near the continuum level.
We used the cross-correlation function (CCF) of the \textsc{HermesDRS} data
reduction pipeline to calculate bisectors (see Fig.~\ref{plot7}). Our CCF is
typically based on $\sim$1650 spectral lines for each spectrum and this
results in  a very good accuracy for the bisector measurement. We
are interested in relative and not absolute bisector measurements and according to
\citet{Martinez05}, the use of the average of many lines is appropriate for studying variations of line bisectors with
time (see also e.g. \citet{Nowak13}). 

The best way to detect bisector changes is to calculate the bisector velocity span
(BVS), which is the difference between bisectors at two different flux levels of a spectral
line. As in \citet{Hatzes15}, we measured the BVS of the profile using the difference of the average
bisector values between flux levels of 0.1 -- 0.3 and 0.7 -- 0.9 of the
continuum value, therefore avoiding the spectral core and wing, where errors of the bisector
measurements are larger. To estimate errors of bisectors we have used a
formula of \citet{Martinez05}, who modified the expression given by
\citet{Gray83,Gray88}. Errors of our BVS
measurements were then calculated by adding quadratically the errors for the
top and bottom zones of the bisector measurements. Both BVS and their errors are
listed in Table~\ref{table1}.

\begin{figure}
\rotatebox{270}{\includegraphics[width=66mm]{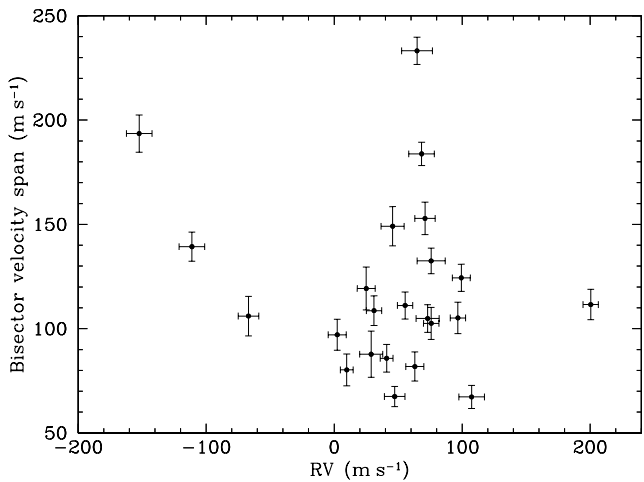}}
\caption{The BVS versus RV measurements for HD~175370 observed by the HERMES
spectrograph.} 
\label{plot8}
\end{figure}

The BVS measurements are shown in the second panel from the bottom of
Fig.~\ref{plot5}. 
The BVS variations as a function of RV are shown in Fig.~\ref{plot8}. 
Pearson's correlation coefficient between RV and BVS variations amounts to -0.21
and the probability that they are uncorrelated is 0.29, indicating that they
are not correlated. However, there are only a few BVS measurements
around the minimum and maximum RV values and more data around these extremes
would be useful. 

We searched for periods in the BVS variations using the program
\textsc{Period04} \citep{Lenz05} and did not find any significant periods. The frequency
spectrum is displayed in the top panel of Fig.~\ref{plot6}.  

\subsection{Chromospheric activity}\label{activity3}

The equivalent width (EW) variations of the \ion{Ca}{ii} H \& K lines are often
used as a chromospheric activity indicator because they are sensitive to stellar
activity which in turn may affect the measured RV variations. In
chromospherically active stars, the \ion{Ca}{ii} H \& K lines show a typical line
core reversal \citep{Pasquini88}. Mercator spectra of HD~175370 have a low S/N 
in the region of the \ion{Ca}{ii} H \& K lines and it is difficult to see if any 
emission features are present or not in the \ion{Ca}{ii} H \& K line cores. 

Instead, we chose to measure chromospheric activity using the \ion{Ca}{ii} triplet lines
\citep{Larson93, Hatzes03, Lee13}. Among the triplet lines, \ion{Ca}{ii} 8662 \AA~
is not contaminated by atmospheric lines near the core. Additionally, 
Mercator spectra are contaminated by a light leak from the calibration 
fibre into the science fibre as displayed on the CCD in spectral regions
containing bright calibration lines, but the region around the \ion{Ca}{ii} 8662
\AA~line is free from the contamination. According to \citet{Larson93}, changes in the core flux of the
line \ion{Ca}{ii} 8662 \AA~are qualitatively related to the variations in the
\ion{Ca}{ii} H \& K lines. We measured the EW at the central part of the
\ion{Ca}{ii} 8662 \AA~line from
8661.0 to 8663.27 \AA~and also from 8661.55 to 8662.8 \AA~and did not find any
significant periodic changes. The frequency spectra are displayed in
Fig.~\ref{plot6} in the second panel from the top. The variations of the
\ion{Ca}{ii} 8662 \AA~line are
very small, which is typical for an inactive star.

\section{Conclusions and discussion}\label{conclusions}

Our spectral analysis of HD~175370 corresponds to a K2$\,$III giant star. 
No carbon depletion or oxygen enhancement was detected in high-resolution
spectra indicating that HD~175370 is a low-mass star, where hydrogen burning 
occurred in a radiative core, dominated by the proton-proton reactions so that 
the CNO cycle did not play a crucial role \citep{Charbonnel93}. 
Based on the \textsc{GBM} code \citep{hekker2013}, we
deduce that HD~175370 is most likely a RGB or AGB star, and not a RC star. 

We conclude that the RV variations of HD~175370 are caused by a low-mass stellar
companion with an orbital period of $\sim 88$ years in a highly eccentric orbit and 
a possible planetary companion with
an orbital period $P=349.5\pm 4.5\jed{d}$, eccentricity
$e=0.22\pm 0.10$, and
a semi-amplitude $K=133\pm 25\jed{m\,s^{-1}}$. Our
interpretation of the RV changes in terms of a planet is supported by the lack
of variability in the spectral line bisectors, \textit{Kepler} light curve, and
the lack of chromospheric activity corresponding to the orbital period of
the planet. 
Furthermore, if 349.5 d 
were the rotation period of the star then this would result
in a {\it maximum} projected stellar rotational velocity of 3.4 $\pm$ 0.5
$\rm{km\,s^{-1}}$ (RGB star) and 3.5 $\pm$ 0.5 $\rm{km\,s^{-1}}$ (AGB star),
or nearly a factor of two smaller than our measured $v\,\sin i$.

However, even the lack of stellar variability in stellar activity indicators 
corresponding to the orbital period of the planet do not prove its existence.
One caveat is 42~Dra which
has been claimed to host a planetary companion \citep{Doellinger09}. 
Our continued RV measurements of 42~Dra show that in the past three years
the amplitude of the orbital motion decreased by a factor of four
(Hatzes et al., in preparation). This casts serious doubts on the planet
hypothesis for the RV variations. When the RV amplitude was high there 
were no variations in H$\alpha$, the spectral line bisectors, or in the
\textit{Hipparcos} photometry. In this case the standard tools for planet
confirmation - the same ones that we use for HD 175370 - failed.
We suggest that measurements of such standard activity indicators 
may not be sufficient and that long-term monitoring is essential in order 
to confirm planets around K-giant stars.

We note that we do not see any variations in the RV amplitude of the
orbital motion due to HD~175370~b. The star 42~Dra showed coherent RV
variations with a constant amplitude over three years. The RV variations
in HD~175370 are constant for more than five years.

We found two periods in the \textit{Kepler}
light curves of 389.4~d and 370~d. Both of them are very close to the orbital
period of the satellite, 372.5~d. We have shown that most of the \textit{Kepler}
giant stars from our sample show the dominant period of $390\pm
20\jed{d}$, which agrees with the orbital period of the \textit{Kepler}
satellite within the 1-$\sigma$ error bars. We
therefore believe that finding a period in the \textit{Kepler} photometry
which is close to the orbital period of the satellite is suspicious and
should not be related to a real physical process unless one looks at all K
giants in the field and detrends them systematically. Even though we cannot
prove that the RV period is unrelated to the photometric periods, there is
no evidence about the opposite. Continued RV monitoring of this system 
may help answer this uncertainty.

A possible explanation of the RV changes observed for HD~175370 could be
an unknown envelope pulsation causing
photometric and spurious Keplerian-like RV variations.
\citet{Jorissen16a,Jorissen16b} found among 13 low-metallicity giants three cases
where small amplitude variations ($K$ ranging from 0.1 to 0.9
$\rm{km\,s^{-1}}$) with periods very close to 1 year are superimposed on a long-period Keplerian
orbit. They do not give a conclusive answer on the origin of these
variations, but leave an envelope pulsation as an option. In choosing a
hypothesis based on a known phenomenon (exoplanets) as opposed to an unknown
phenomenon (envelope oscillations) we chose the former until more evidence
comes to light. In addition, we have shown that the
photometric periods we have found in the \textit{Kepler} light curves of
HD~175370 are suspicious of being due to the orbital period of the satellite.
   
Oscillation modes dominate the short time-scale variations in the light curve and  
their amplitudes are at least ten times larger than the expected transit
depth. It would therefore be difficult to detect any transit events in
the light curve, which would constrain the orbital inclination. 

HD~175370 is one of the few close binary systems to host a giant
planetary candidate and thus may be important for understanding planet formation in
binary systems. To date, only about six binary systems with separations
less than 25~AU have been found to host giant planets
\citep{Morais08,Ramm09}. With a binary orbital separation of 22~AU 
HD~175370 adds to this small list. HD~175370 is unique in that it has the largest binary
eccentricity ($e$ = 0.88) of these binary systems hosting planetary
candidates. At closest approach the primary-secondary separation is only
a factor of 2.7 larger than the planet-star semi-major
axis. 
However, we fixed the orbital period and eccentricity of the binary
orbit to values obtained from varying these parameters separately and searched for
a minimum in the residuals, and therefore the errors of both parameters are
expected to be large, which also implies a large error on the orbital
separation. We estimate that uncertainties of the period and eccentricity 
are $\sim$4000 d and $\sim$0.02, respectively, which implies that at closest approach 
the primary-secondary separation
could be a factor of 3.5 larger than the planet-star semi-major axis.
A dynamical study may place additional constraints
on the age and future evolution of this system.

HD~175370~b is one of the first planetary candidates 
discovered around a \textit{Kepler} giant star via the RV method. 
Unlike for other evolved giant stars with extrasolar planets detected by ground-based RV
surveys, for HD~175370 we have high-quality \textit{Kepler} photometry from
which we can determine the stellar mass and radius in a more or less model independent
way. The relative uncertainties of our derived stellar mass, stellar radius, and $\log g$
are 16, 14, and 1 per cent, respectively.
HD~175370 is a relatively old star with an age of at least 5.4 Gyrs. More discoveries of planets around 
evolved giant stars can give us a better understanding of the mass dependence of planet
formation and the evolution of planetary systems.

\section*{Acknowledgements}

The data presented here have been taken using the 2-m Alfred Jensch 
Telescope of the Th\"uringer Landessternwarte Tautenburg. We are grateful to
the personnel of the observatory for their support during our observations.
M.Hr. acknowledges the support of the Deutsche Forschungsgemeinschaft (DFG) 
grant HA 3279/5-1. The research leading to the presented results has received 
funding from the European Research Council under the European Community's Seventh Framework
Programme (FP7/2007-2013)/ERC grant agreement no 338251 (StellarAges). 
\'A.S. acknowledges financial support of the Hungarian National
Research, Development and Innovation Office -- NKFIH K-115709 and OTKA
K-113117 grants, and the J\'anos Bolyai Research Scholarship of the
Hungarian Academy of Sciences. We thank David F. Gray for his kind gesture of giving his
book on spectral-line analysis and Eike W. Guenther for establishing the 
data reduction procedure for TLS data. This research has made use of the electronic bibliography 
maintained by NASA-ADS system and the SIMBAD database, operated at CDS, Strasbourg, France.

%%%%%%%%%%%%%%%%%%%%%%%%%%%%%%%%%%%%%%%%%%%%%%%%%%



%\bibliographystyle{mnras}


%%%%%%%%%%%%%%%%%%%%%%%%%%%%%%%%%%%%%%%%%%%%%%%%%%

%%%%%%%%%%%%%%%%% APPENDICES %%%%%%%%%%%%%%%%%%%%%

%\appendix


%%%%%%%%%%%%%%%%%%%%%%%%%%%%%%%%%%%%%%%%%%%%%%%%%%

% Don't change these lines
\bsp	% typesetting comment
\label{lastpage}
\end{document}